\documentstyle[prl,aps]{revtex}

\input{psfig.sty}

\begin{document}
\draft

\twocolumn[
\hsize\textwidth\columnwidth\hsize\csname @twocolumnfalse\endcsname
\title{L\'{e}vy flights in quantum transport in quasi-ballistic wires. }

\author{M. Leadbeater$^a$, V.I. Falko$^{b,c}$ and C.J. 
Lambert$^b$} \address{
$^{a}$ Max-Planck-Institut f\"ur Physik komplexer Systeme, 
N\"othnitzer Str. 38, 01187 Dresden, Germany\\
$^{b}$ School of Physics and Chemistry,
Lancaster University, Lancaster, LA1 4YB, UK\\
$^{c}$ Departement de Physique, Universit\'{e} Joseph Fourier, 
Grenoble 1, France}

\date{\today}
\maketitle

\begin{abstract}

Conductance fluctuations, localization and statistics of Lyapunov
exponents are studied numerically in pure metallic wires with rough 
boundaries (quasi-ballistic wires). We find that the correlation energy
of conductance fluctuations scales anomalously with the sample
dimensions, indicating the role of L\'{e}vy flights. Application of a 
magnetic field deflects the 
L\'{e}vy flights which reduces the localization length. 
This deflection also breaks the geometrical
flux cancellation and restores the usual Aharonov-Bohm type
magneto-conductance fluctuations.

\end{abstract}
\narrowtext]

During recent years theoretical interest in mesoscopic systems has started
to divert from studies of universal properties of ergodic systems, 
such as disordered metals, towards structures which are non-ergodic 
\cite{AltRev}. Ergodicity of diffusive conductors is provided by effective
mixing of the phase space corresponding to the classical counterpart of
quantum mechanics - due to quantum scattering at a short-range random
potential. In non-ergodic systems, homogeneous mixing of classical phase space 
does not occur before the quantum state is formed. Although issues of
non-ergodicity are usually addressed in a context of spectral and wave
function statistics in closed systems, such as partly chaotic 
billiards \cite{Berry,Gutzwiller,Andreev}, one may extend the notion of
non-ergodicity to open systems and study its implications in quantum
transport effects such as localization and conductance
fluctuations. We distinguish below a class of disordered objects
where a large part of classical phase space is not mixed \cite{Aleiner}
since the particle motion through it is not diffusive. These systems 
are pure metallic wires with rough edges corrugated at the
length scale of the electron wavelength (quasi-ballistic wires) \cite
{McGurn,Tesanovich,DugaevKhm,Been,VF,MacKinnon,Harris,Yurkevich}, where the
existence of anomalously long grazing ballistic paths is known to lead to
noticeable effects even in classical transport properties \cite{Pippard}.

In the theory of dynamically generated random walks, ballistic paths
with lengths very much exceeding the average have the name L\'{e}vy
flights \cite{LevyFlightReview,LevyFlBook} and when the latter occur with a
relatively high probability they determine a superdiffusion character of
random motion. In diffusive wires with static short-range bulk
impurities, the probability of finding a ballistic path $\eta $ longer than the
mean free path, $l$ is exponentially small, $\sim \exp (-\eta /l)$. This
leads to the normal diffusion relation between the variance of the lengths
of random walk paths and the system size and gives 
the 'ergodic' form for the correlation energy of
conductance fluctuations, $E_{{\rm c}}\sim hD/L^{2}$ \cite{AltRev}. A
natural example of a classically superdiffusive system with static
disorder is a pure metallic wire with corrugated edges, where L\'{e}vy
flights are just those grazing ballistic trajectories which cross it at a
small angle, $\alpha \ll 1$, to the wire axis. In wires with only edge
disorder, ballistic segments $\eta =W/\sin \alpha $ much longer than the
sample width appear with the probability $\sim \eta ^{-2}$ for $\eta \gg W$.
Below we report the results of numerical simulations of quantum transport
in quasi-ballistic wires with edges which are rough on an atomic
scale. We analyze the effects of superdiffusion on the correlation
properties of mesoscopic conductance fluctuations, $\delta G(\epsilon ,B)$ and 
the effect of a magnetic field on the localization length, $L_{%
{\rm c}}$. To anticipate a little, the correlation energy of conductance
fluctuations we found scales anomalously with the system size as $E_{{\rm c}%
}=\frac{1}{2}hWv_{F}L^{-2}\ln (\frac{L}{1.7W})$. Although geometrical
flux cancellation \cite{DugaevKhm,Been,VF} hinders the conventional
crossover in the localization effects between the orthogonal and unitary
symmetry classes \cite{Efetov,Pichard}, we observe the usual Aharonov-Bohm
magnetoconductance fluctuations due to the deflection of L\'{e}vy flights by
a magnetic field (which transforms part of them into skipping orbits). This
deflection breaks the flux cancellation rules and also determines the
magnetic field dependence of the localization length which is different to that
obtained in ergodic (with bulk impurities) disordered systems.

The results presented below are based upon numerical solution of a
two-dimensional Anderson Hamiltonian on a square lattice, $%
H=\sum_{i}|i\rangle \epsilon _{i}\langle i|-V\sum_{\langle ij\rangle
}|i\rangle \langle j|$, where $\langle ij\rangle $ denotes nearest neighbour
sites $i$ and $j$. The structure considered consists of two ideal leads of
width $W$ attached to a scattering region $W$ sites wide and $L$ sites long
(all lengths are in units of the lattice constant $a$). In the absence of a
magnetic field, the off-diagonal matrix elements $V=1$ determine the width
of the energy band. Within the leads and in the centre of the scattering
region (when modeling clean wires with rough boundaries), the diagonal
matrix elements were $\epsilon _{i}=\epsilon _{0}$ (with $\epsilon _{0}>1$,
which keeps the Fermi level away from the van Hove singularity in the band
center). For simulating bulk disorder, the $\epsilon _{i}$ in the scattering
region are taken uniformly from the interval $-U/2<\epsilon _{i}-\epsilon
_{0}<U/2$ where $U$ is the disorder strength. For sites on the boundary, $%
\epsilon _{i}=\epsilon _{0}+\epsilon _{B}$ where $\epsilon _{B}=10^{4}$. The
rough structure of the boundary was generated by having an equal probability
of either 0, 1 or 2 sites at each edge with on-site potential $\epsilon
_{0}+\epsilon _{B}$ \cite{RemarkWidth}. To simulate the effect of a magnetic
field, we incorporate a Peierls phase factor into $V$ in the scattering
region. Numerical decimation techniques \cite{Lambertetc} were used to
compute the Greens function for each given realization of disorder. From
these the transmission matrix $tt^{\dagger }$ was found and diagonalized.
Its eigenvalues, $T_{n}$, were analyzed statistically and used to compute
the conductance $G\,=\,\sum_{n}T_{n}$, which we measure in units of ${\frac{%
e^{2}}{h}}$. Note that numerical results presented here describe a model
system without any decoherence ({\it i.e.}, without any inelasticity).

The effect of L\'{e}vy flights can be easily identified in our numerical
results of quantum transport in pure wires with edge disorder. Rather than
consider the statistics of the $T_{n}$'s it is more useful to introduce the
parametrization $T_{n}=1/\cosh ^{2}(L/\xi _{n})$ and consider the
distribution, $P(\xi ^{-1})$, of inverse localization lengths or Lyapunov
exponents. This is shown in Fig. 1(a) for 4 series of quasi-ballistic wire
samples with $W=15$ (giving a mean width of $13$) and lengths $L=52$ (A), $%
104$ (B), $208$ (C) and $416$ (D), and a series of samples (U) with bulk
'defects'. As pointed out by Tesanovich {\it et al} \cite{Tesanovich} and
verified numerically in Ref. \cite{MacKinnon}, the length of L\'{e}vy
flights, $\eta _{\max }$, is limited in quantum systems. This is because
uncertainty in the transverse momentum, $\delta k_{\perp }\sim \hbar /W$, in a
wire with a finite width sets a quantum limit to the angles $W/\eta \sim
\alpha >\delta \alpha \sim \delta k_{\perp }/k_{F}\sim \lambda /W$ which can
be assigned to a classically defined ballistic segment \cite{VF}. This sets
the cut-off $\eta _{\max }=W^{2}/\lambda _{F}$, and a finite localization
length $L_{c}\sim \frac{\pi W^{2}}{\lambda _{F}}$. Samples from the series A
and B meet the criterion $L<L_{c}$ and the distribution of the Lyapunov
exponents obtained in them has a pronounced peak at small $\xi ^{-1}$
corresponding to eigenvalues $T_{n}\approx 1$. This should be compared to
the plateaux-like \cite{Chalker} distribution $P(\xi ^{-1})$ reproduced
using the same numerical procedure in a sample from the series U. \ Note a
finite width of the ballistic peak in $P(\xi ^{-1})$ and that $P(0)=0$. The
enhanced density of small $\xi ^{-1}$ can be identified even in samples from
the series C and D with $L\approx L_{{\rm c}}$, where $P(\xi ^{-1})$ starts
to show a periodic modulation specific to the localized regime where the
spectrum of Lyapunov exponents tends to crystallize \cite{Pichard,Frahm1}.

The distribution of the eigenvalues of the transmission matrix results in a
finite-width distribution of conductances and, hence, conductance
fluctuations. The statistics of conductance fluctuations, $P(G)$, for wires
with edge disorder is shown in Fig. 1(b) (for the case of zero magnetic
field) for the series A, $W=15$, $L=52<L_{{\rm c}}\sim \pi W^{2}/\lambda
_{F}\sim 200$, $\langle G\rangle \approx 2.0$ (the distribution is almost
Gaussian) and C, $W=15$, $L=208\sim L_{{\rm c}}$, $\langle G\rangle \approx
\,0.71$. The distribution function $P(G)$ is the result of the analysis of
various realizations of disorder. We shall denote such averaging over
realizations by $\langle ....\rangle $. When calculating correlation
functions, we also add averaging over energy. The sample length dependence
of the correlation energy, $E_{{\rm c}}(L)$, of the fluctuation pattern
found from the half-height of the correlation function $\langle \delta
G(\epsilon )\delta G(\epsilon +\delta \epsilon )\rangle $ is shown in Fig.
1(c). For comparison, $E_{{\rm c}}(L)\sim L^{-2}$ is also shown for
bulk-disordered samples (U), which serves us as a reference case.

In wires with edge disorder ballistic segments $%
\eta =W/\sin \alpha $ much longer than the sample width appear with the
probability $\pi ^{-1}|\frac{d\eta (\alpha )}{d\alpha }|^{-1}\sim \eta ^{-2}$
for $L>\eta \gg W\gg \lambda _{F}$ and this gives rise to a 
difference of the correlation energy sample length dependence from the
aforementioned 
ergodic law. The correlation energy is 
determined by the lowest eigenvalue in the
relaxation spectrum of the diffusion operator. In the anomalous case, the
diffusion process is described by the integral equation 
\[
P\left( \frac{t}{\tau };x,x^{\prime }\right) =\frac{1}{2}\int_{-\frac{L}{2W}%
}^{\frac{L}{2W}}\frac{P\left( \frac{t}{\tau }-\sqrt{1+(x-y)^{2}};y,x^{\prime
}\right) }{\left[ 1+(x-y)^{2}\right] ^{3/2}}dy
\]
with initial conditions $P(t=0;x,x^{\prime })=\delta (x-x^{\prime })$. The
correlation function $P(t/\tau ;x,x^{\prime })$ describes the probability
density to find a particle at the diffusively scattering edge of a sample
symmetrized over both wire edges, and it is represented in the coordinates
along the wire axis normalized by the wire width. The time is normalized by
the ballistic time, $\tau =W/v_{F}$. The solution of this equation in a long
system can be found by taking the Fourier transform, 
\[
P(\omega ,q)=\frac{1}{\Pi }=\left[ 1-\int_{-\infty }^{\infty }\frac{\exp
\left( iqx+i\omega \sqrt{1+x^{2}}\right) }{2\left[ 1+x^{2}\right] ^{3/2}}dx%
\right] ^{-1}.
\]
To study the low-lying spectrum of $P(\omega ,q)$, we expand $\Pi $ over $%
\omega \ll 1$, and find that, 
\[
\Pi \approx (1-qK_{1}(q))-\frac{i\pi \omega }{2}+\frac{\omega ^{2}K_{0}(q)}{2%
}+...
\]
where $K_{n}$ are Bessel functions. In the limit of $q\ll 1$ this gives 
\begin{equation}
\Pi \approx -i\frac{\pi }{2}\omega +\frac{1}{2}q^{2}\ln \left( \frac{%
2e^{1/2-\gamma }}{q}\right) .  \label{eq1}
\end{equation}
For a sample of length $L$, we take $q_{\min }=\pi W/L$ and get 
\begin{equation}
E_{{\rm c}}\sim \frac{hv_{F}W\ln (L/cW)}{2L^{2}},\,\,\,c=\frac{\pi }{2}%
e^{\gamma -1/2}\approx 1.7,
\end{equation}
where $\gamma $ is Euler's constant. The anomalous dependence in this
analytical result describes a faster escape of a particle from a
quasi-ballistic wire, as compared to a bulk-diffusive one, and it is
represented by the solid curve in Fig. 1(c). Note that to relate the above
semiclassical analysis to purely quantum numerical simulations, we use an
effective sample width $\langle W\rangle -\lambda _{F}$. In addition, it is
possible to calculate the variance of conductance fluctuations in a
quasi-ballistic case, $\langle \delta G_{{\rm qb}}^{2}\rangle $. This has
been done by evaluating the probabilities of two paths to encounter each
other at two scatterers, which uses the result of Eq. (\ref{eq1}). Such a
calculation shows that $\langle \delta G_{{\rm qb}}^{2}\rangle /\langle
\delta G_{{\rm erg}}^{2}\rangle =\frac{64}{\pi ^{4}}\approx 0.66$, whereas
the numerical result obtained for a system with $W=15$ and $L=104$ is $%
\langle \delta G_{{\rm qb}}^{2}\rangle /\langle \delta G_{{\rm erg}%
}^{2}\rangle \approx 0.72$.

Another feature of quasi-ballistic wires is flux cancellation \cite
{DugaevKhm,Been,VF}, which is an exact geometrical fact in ballistic systems
with parallel edges. Because of this cancellation, no Aharonov-Bohm flux can
be encircled by loops composed of a sequence of ballistic flights between
sample boundaries. In the metallic regime, $L<L_{{\rm c}}$, the curving of
an electronic trajectory by a magnetic field reduces the length of the
longest ballistic paths, so that the conductance $\langle G\rangle $
declines when the transverse deflection of the L\'{e}vy flights, $\delta
x_{\perp }\approx \frac{1}{2}(ev_{F}B/mc)\left( L/v_{F}\right) ^{2}\sim
BL^{2}\lambda _{F}/\phi _{0}$ becomes larger than the wire width, $\delta
x_{\perp }\geq W$: 
\begin{equation}
{\rm at}\;\;B>B_{{\rm defl}}\sim \phi _{0}W/(\lambda _{F}L^{2})\sim (L_{{\rm %
c}}/L)(WL)^{-1}\phi _{0}.
\end{equation}
This effect is present both in the averaged (over disorder) conductance for
samples from series A and B shown in Fig. 2(a) and in the suppression of a
peak at $\xi ^{-1}$ in $P(\xi ^{-1})$ in Fig. 2(b) for samples from the same
series, as in Fig. 1(a). The conductance $\langle G\rangle $ decreases up to
the field where the electron cyclotron radius $R_{c}$ becomes commensurable
with the wire width $W$. At such a field, $B_{{\rm com}}=0.55{\frac{\hbar
k_{f}}{eW}}$ \cite{Pippard}, the magnetoconductance has a minimum followed
by a rise and formation of the quantum Hall effect plateaux after which any
electron cyclotron orbit either decouples from the edges, or skips
independently along each of them \cite{RemarkMF}.

The shortening of the length of L\'{e}vy flights reduces the localization
length in a wire. In samples from the series C and D (also shown in Fig.
2(b)), localization appears to be stronger in a field than at $B=0$: The
crystallization in the spectrum of $\xi ^{-1}$ prevails, which reflects the
fact, supported by the analysis of longer wires (up to $L=1040$) , that the
localization length is shortened by a field. This tendency, illustrated in
Fig. 2(c) (solid curve), is opposite to the behavior of $L_{{\rm c}}(B)$ 
\cite{Efetov,Pichard} in diffusive samples from the U-series (dashed curve).
The relevant scale of magnetic fields at which the tendency of the
localization length to shorten in quasi-ballistic wires starts, $\phi
_{0}\lambda _{F}/W^{3}$, can be obtained from $B_{{\rm defl}}$ by replacing
the sample length $L$ with $L_{{\rm c}}$, which formally gives a similar
scale as a field which would provide the Aharonov-Bohm type of a crossover
in $L_{{\rm c}}$ between two symmetry classes, so that the latter is
hindered by the deflection effect.

The same deflection of L\'{e}vy flights at $B>B_{{\rm defl}}$ also breaks
the geometrical flux cancellation and restores the ability of a pair of
electron paths to encircle a magnetic flux $\sim LWB$. At this field the
longest ballistic flights, which still dominate transport, transform into
skipping orbits so that the electron path inside a sample acquires segments
which do not cross the wire: They start and end on the same edge. On the one
hand, the positive magnetoconductance, which could result from the
suppression of the weak localization correction developed at the field scale
of $B\sim B_{{\rm defl}}$, is mostly hindered by a larger negative classical
effect. On the other hand, the break-down of geometrical flux
cancellation manifests itself in pronounced magnetoconductance fluctuations $%
\delta G(B)=G(B)-\langle G\rangle $ illustrated by an example in Fig. 3(a)
(solid line), which shows the conductance $G(B)$ of a single sample, along
with results from a sample from the U-series (dashed line - this has been
shifted by 1 vertically for clarity). The variance of fluctuations is shown
in Fig. 3(b). From analyzing the auto-correlation function as a function of
magnetic field, we deduce the sample length dependence of the correlation
magnetic field $B_{c}$ in quasi-ballistic samples. It corresponds to the
magnetic field flux through the sample area equal to $2.5\phi _{0}$, in
comparison with $1.5\phi _{0}$ that we get in the bulk-disordered case. \
This seems to explain an earlier experimental observation \cite{Houten}.

The authors thank C.W.J. Beenakker for discussions. VF acknowledges
financial support from EPSRC.

\newcommand{\figwidth}{8.0truecm}

\begin{figure}[tbp]
\centerline{\psfig{figure=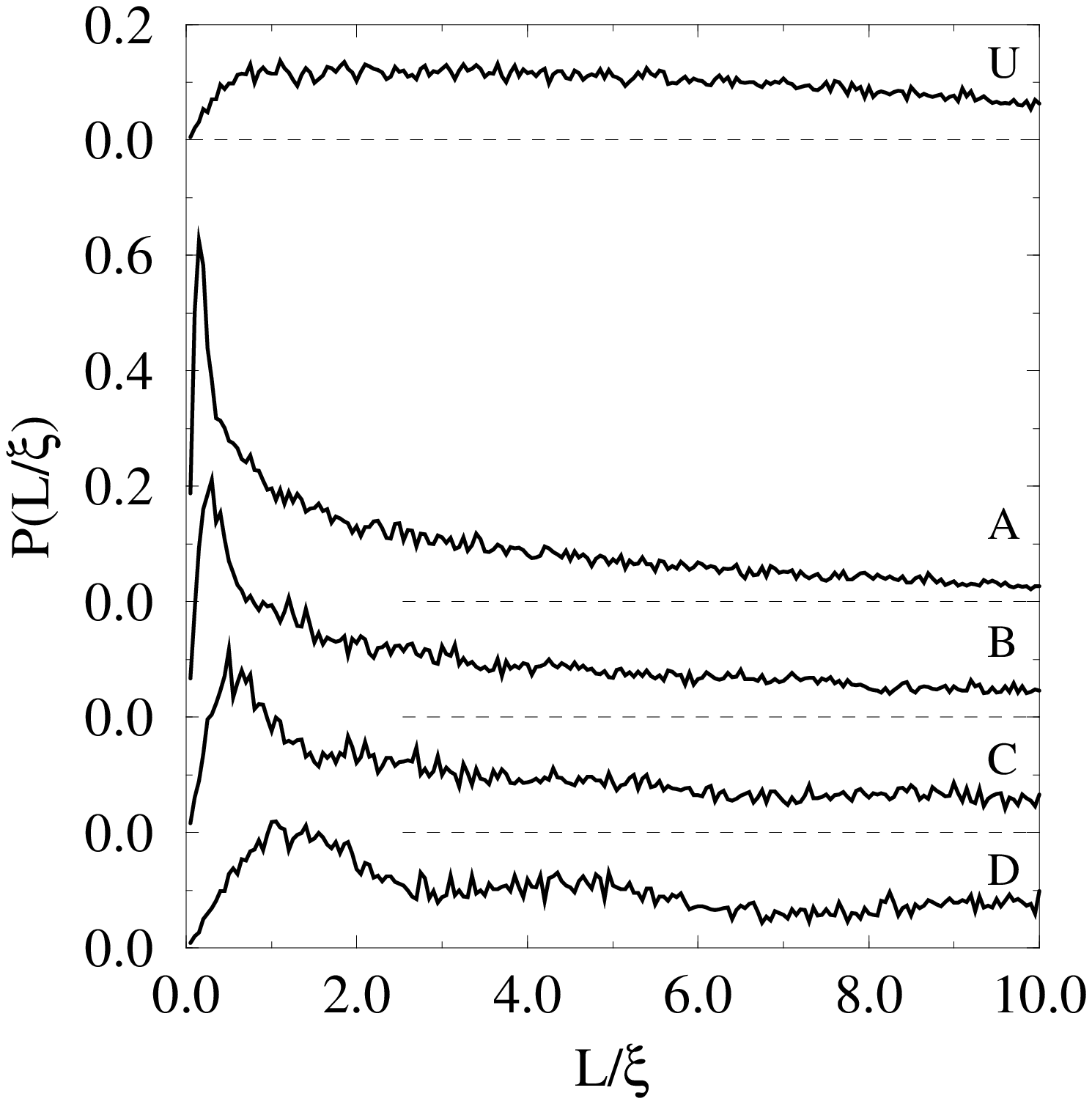,width=\figwidth} }
\vspace*{-0.35truecm}
\centerline{\psfig{figure=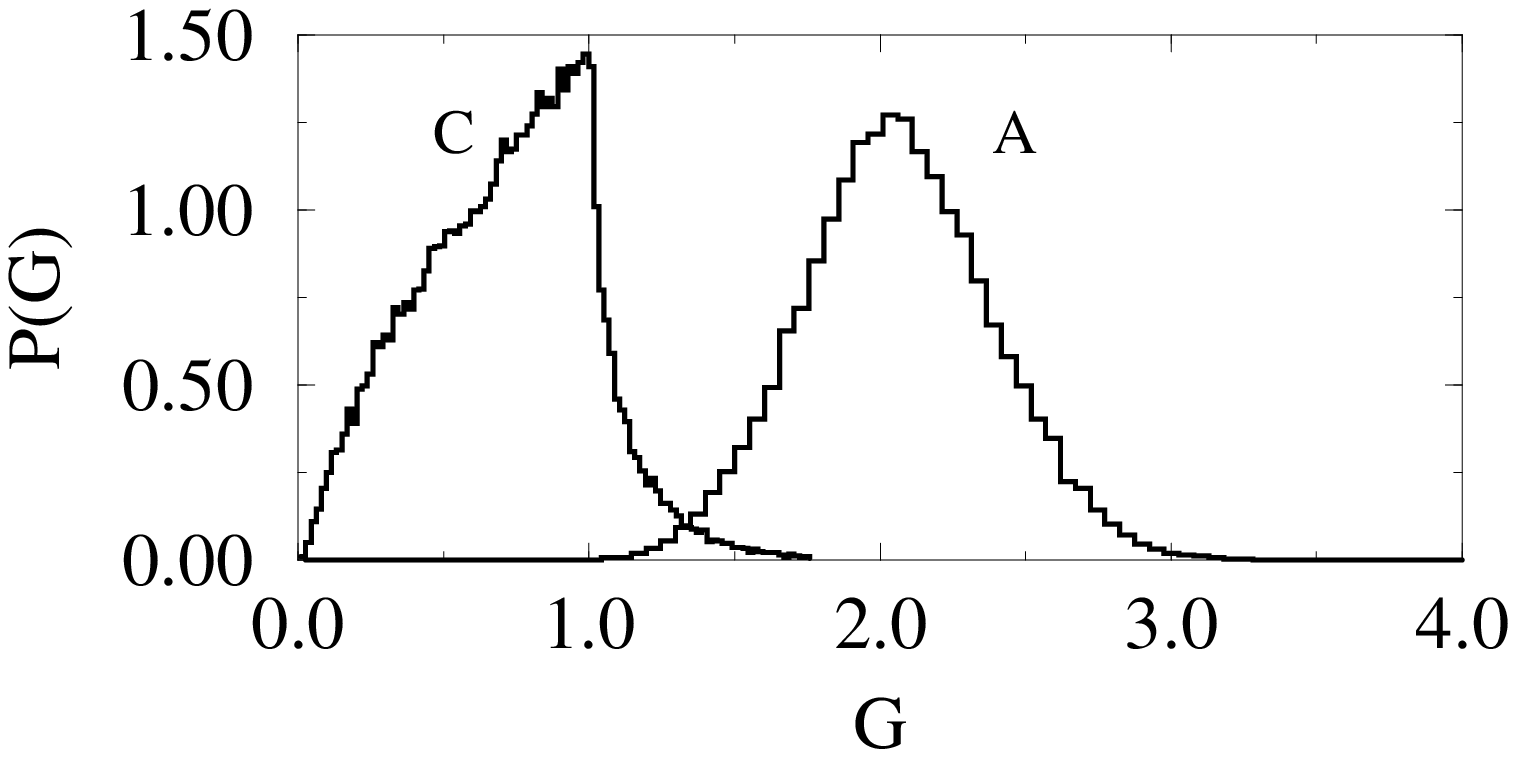,width=\figwidth} }
\vspace*{-0.35truecm}
\centerline{\psfig{figure=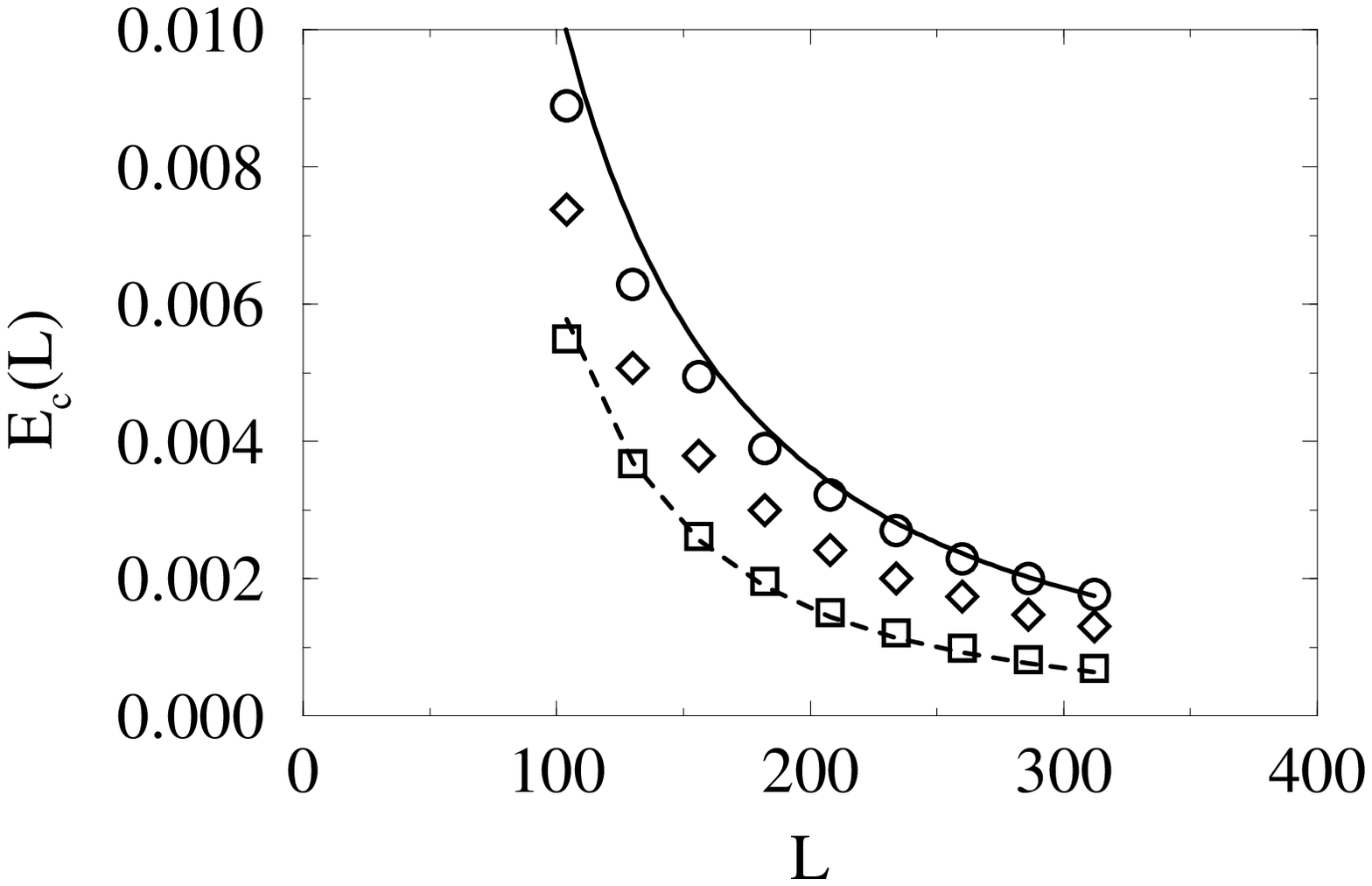,width=\figwidth} }
\caption{(a) Distribution function $P(L/\protect\xi)$ for 4 quasi-ballistic
samples with $W=$15 and $L=$52(A), 104(B), 208(C) and 416(D) and for one
wire (U) with on-site bulk disorder $U=2.0$ (nominal mean free path $%
l\approx 8.5$, $W=15$ and $L= 52$). (b) shows $P(G)$ for the two
quasi-ballistic structures A and C. (c) shows $E_c(L)$ for quasi-ballistic
wires with $\protect\epsilon_0\,\in [1.5,1.7]$ (circles) and $\protect%
\epsilon_0\,\in [1.0,1.2]$ (diamonds) and for a wire from the $U$ series
(squares). Also shown is the analytical result (eq. (2) - solid curve) with $%
hv_f=12.3$ and $W=9.3$ [15]. The dashed curve is the ergodic law $E_{{\rm c}%
}\sim hD/L^{2}$ with $hD=62.5$.}
\end{figure}

\begin{figure}[h]
\centerline{\psfig{figure=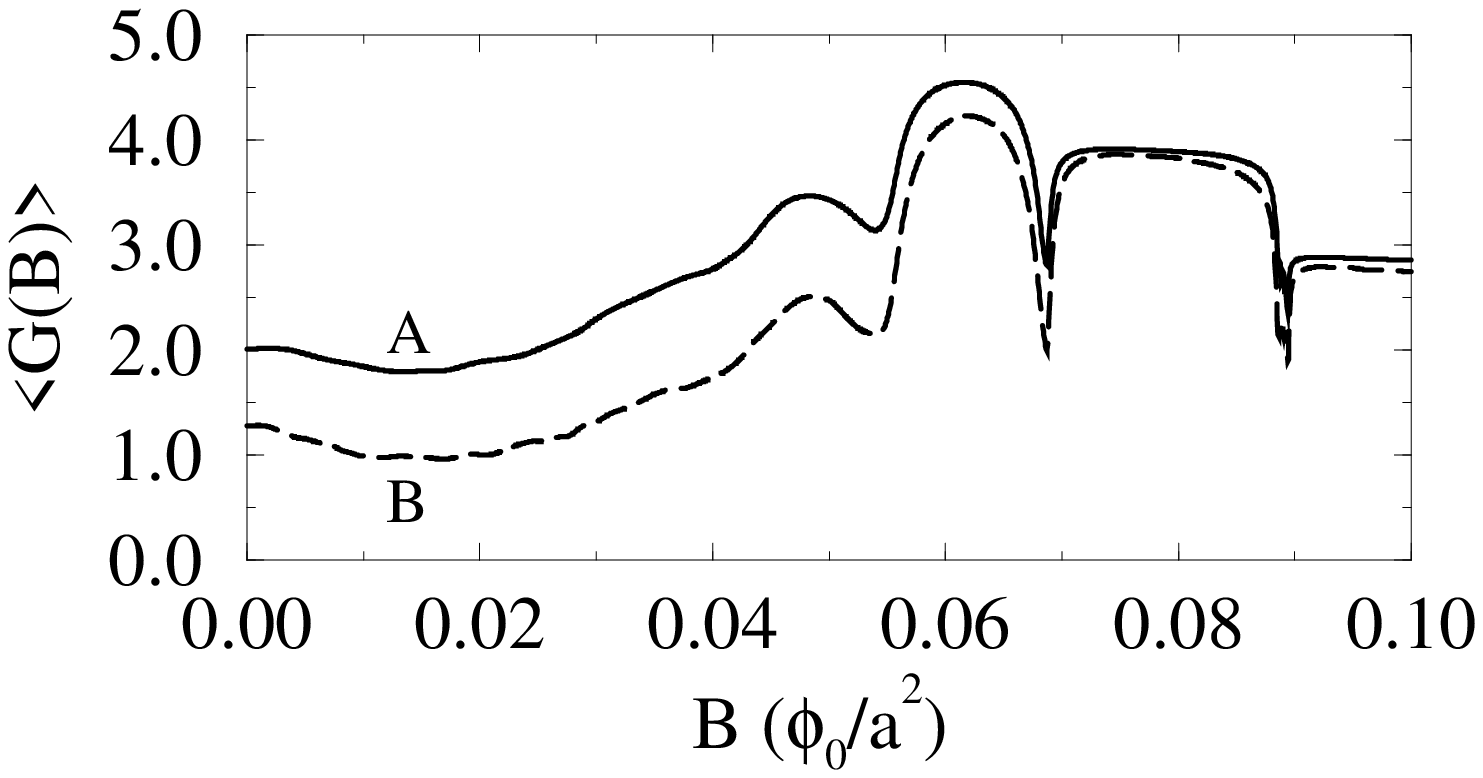,width=\figwidth} }
\vspace*{-0.35truecm}
\centerline{\psfig{figure=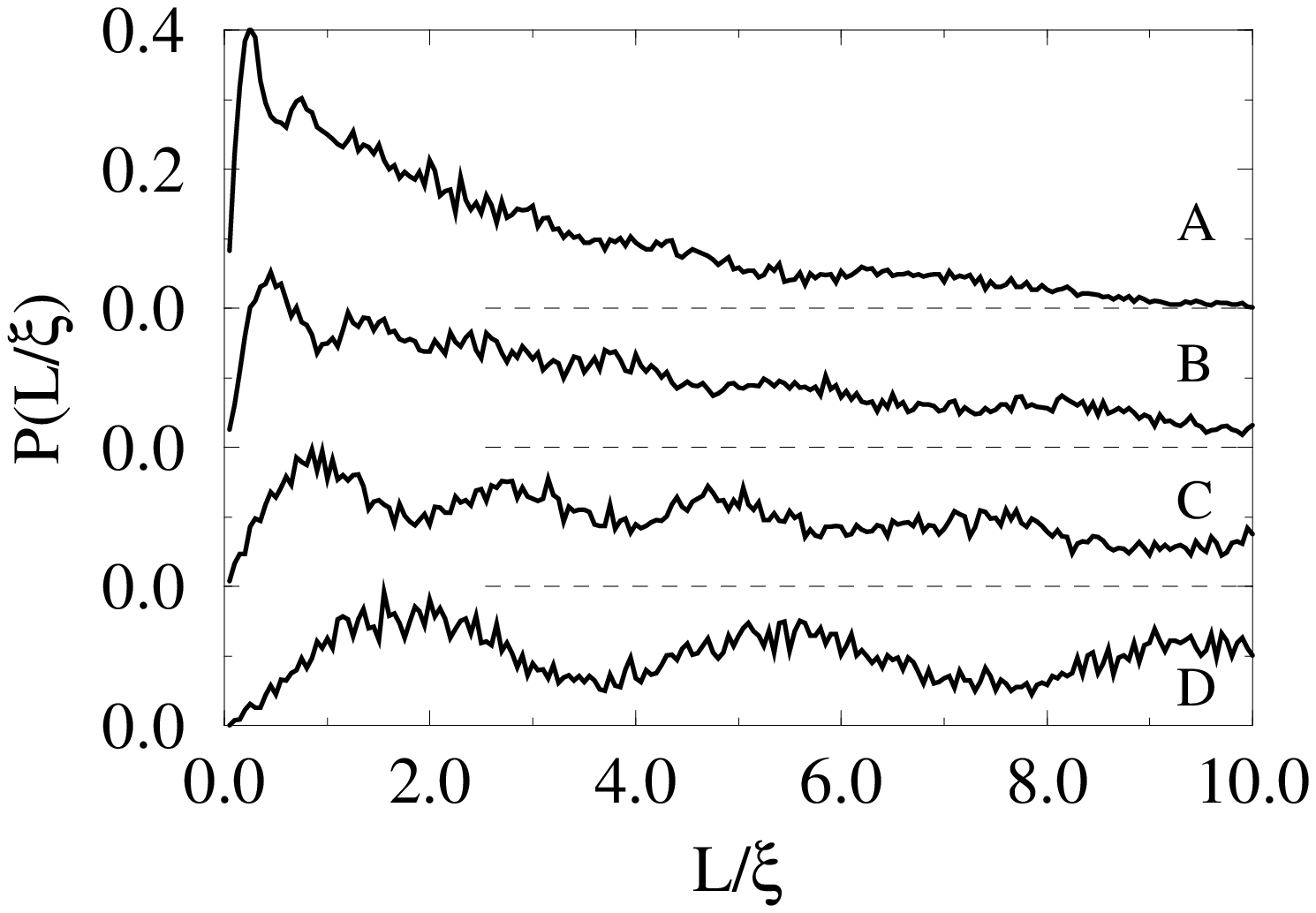,width=\figwidth} }
\vspace*{-0.35truecm}
\centerline{\psfig{figure=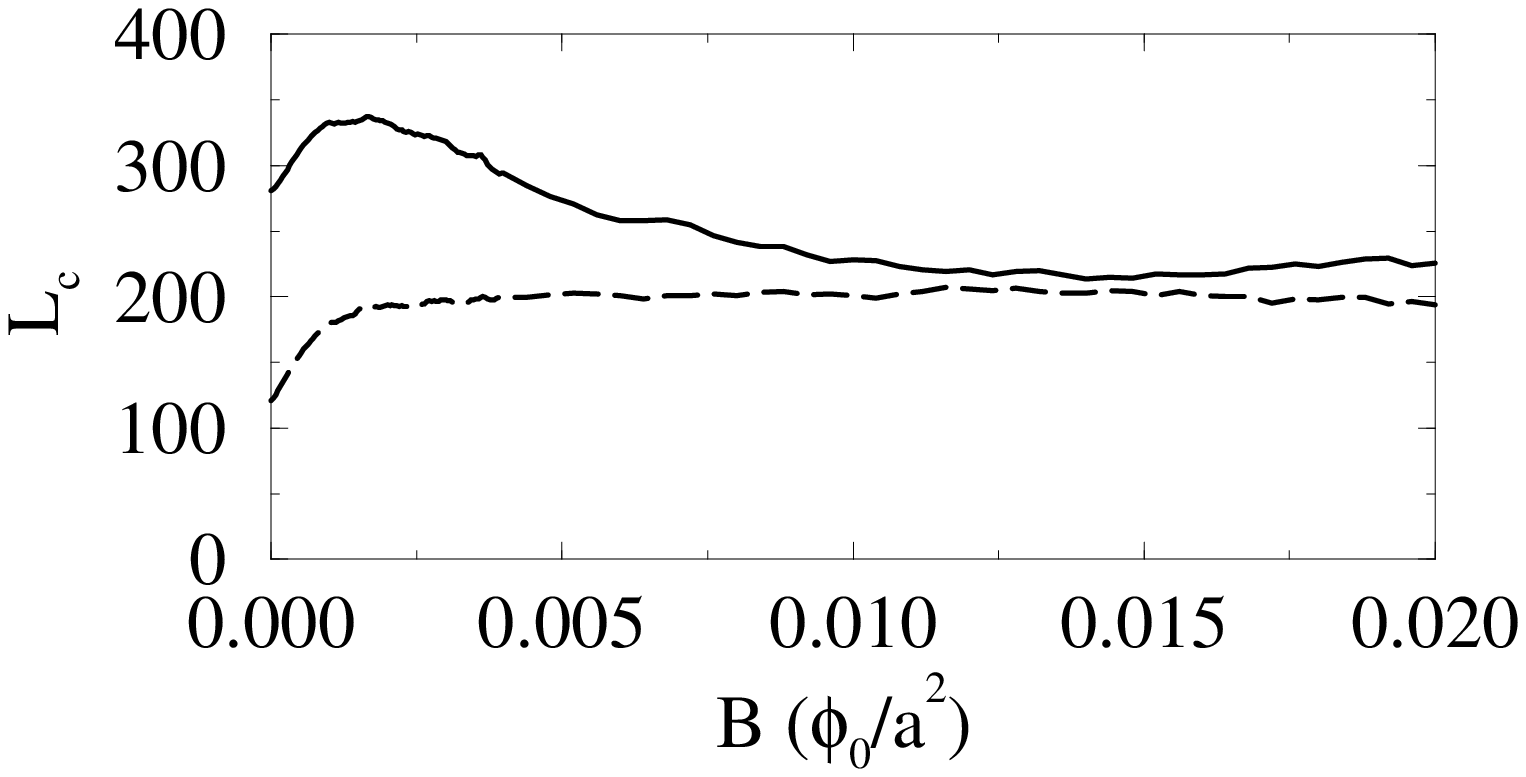,width=\figwidth} }
\caption{(a) The ensemble averaged conductance as a function of a magnetic
field for wires from the A and B series. (b) The same curves (A-D) as for
Fig. 1(b) but at finite magnetic field $B=0.02\protect\phi_0/a^2$. (c) the
localisation length for quasi-ballistic (solid curve) and disordered (dashed
curve - bulk mean free path $l\approx 8.5$) wires.}
\end{figure}

\begin{figure}[tbp]
\centerline{\psfig{figure=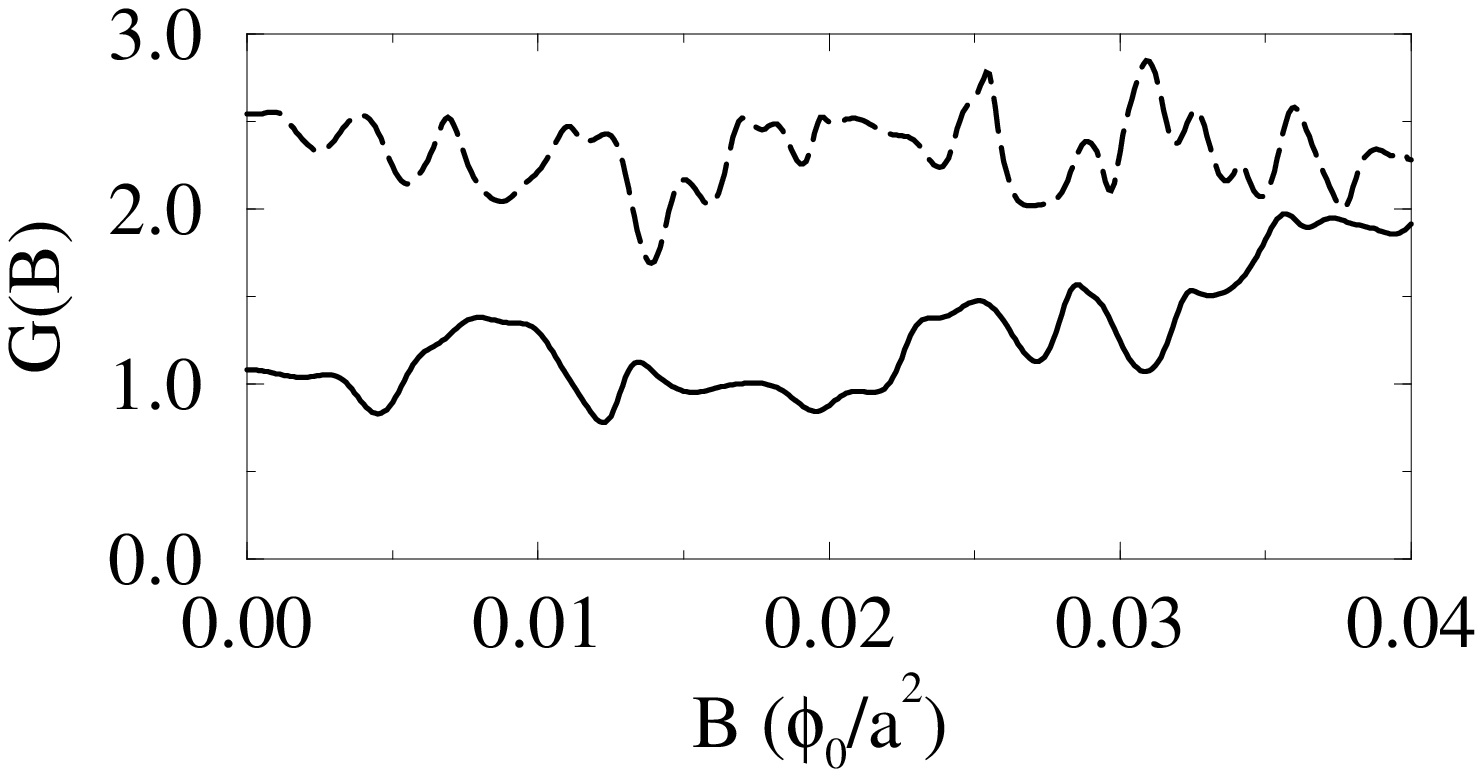,width=\figwidth} }
\vspace*{-0.15truecm}
\centerline{\psfig{figure=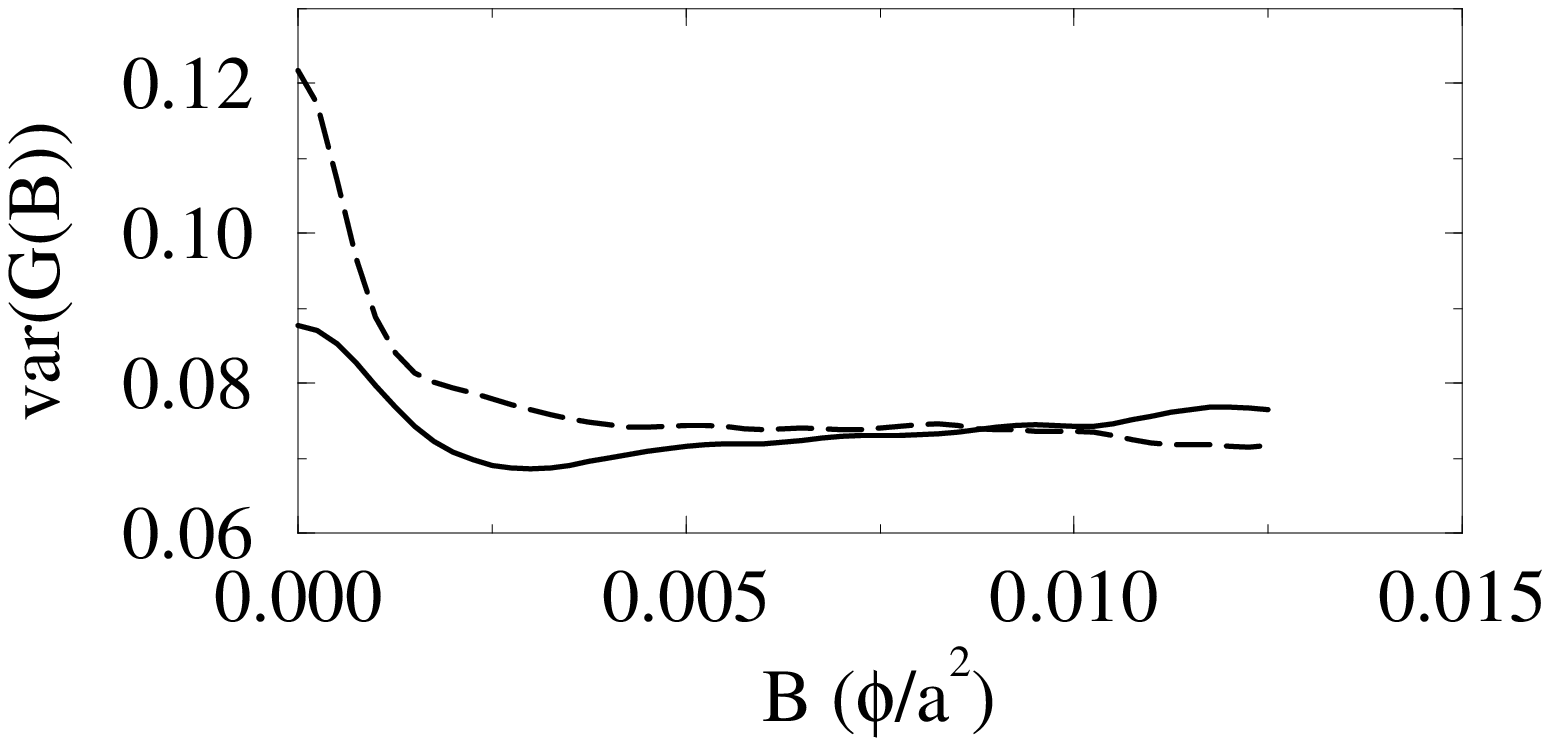,width=\figwidth} }
{3(a) Conductance for one quasi-ballistic wire (solid curve) from the B series
and one disordered wire (dashed curve - mean free path $l\approx 16$).
3(b) Variance var(G(B)) for quasi-ballistic wires from the $B$ series
(solid curve) and disordered wires (dashed curve - mean free path $l\approx 16$)}
\end{figure}


\begin{references}
\bibitem{RemarkWidth}  With this choice the mean width of the scattering
region along its length is $W-2$. Note that the dispersion relation is $%
E\,=\,\epsilon _{0}-2V(\cos (k_{x}a)+\cos (k_{y}a))$ and for L\'{e}vy
flights we take the Fermi velocity to be that along the wire axis (i.e. $%
k_{y}=0$) $\hbar v_{F}=2Va\sin (k_{x}a)$. The choice of $E=0$, $\epsilon
_{0}=1.6V$ leads to $\lambda _{F}\approx 3.7a$ and $\hbar v_{F}=1.96Va$ and
for $E=0$, $\epsilon _{0}\,=\,1V$ we have $\lambda _{F}\approx 3.2a$ and $%
\hbar v_{F}=1.74Va$
\end{references}
\end{document}